\documentclass[]{aipproc}

\input{axodraw.sty}

\layoutstyle{6x9}

\def\myint{\intop\nolimits}

\SetInternalRegister\hbadness{8000} 

\begin{document}

\begin{flushright}
IPAS-PUB-k0011\\
December 2001
\end{flushright}

\title 
      [A Novel PQCD Approach in Charmless B-meson Decays]
      {A Novel PQCD Approach in Charmless B-meson Decays
\footnote{Talk presented by Y.-Y. Keum at the 4th international workshop on
B physics and CP violation, Ise, Japan, Feb. 2001,
and at the 9th international Symposium 
on Heavy Flavor Physics, Caltech, Pasadena, Sept. 10-13,2001}  
}

\classification{43.35.Ei, 78.60.Mq}
\keywords{Document processing, Class file writing, \LaTeXe{}}

\author{Y.-Y. Keum}
{address={Institute of Physics, Academia Sinica, 
Nankang Taipei 151, Taiwan (R.O.C.)},
email={keum@phys.sinica.edu.tw},
}

\iftrue
\author{H-N Li}{
 address={Institute of Physics, Academia Sinica, 
 Nankang Taipei 151, Taiwan (R.O.C.)},
 email={hnli@phys.sinica.edu.tw},
}

\author{A. I. Sanda}{
 address={Department of Physics, Nagoya university, Nagoya 464-01, Japan},
 email={sanda@eken.phys.nagoya-u.ac.jp},
}
\fi

\copyrightyear  {2002}

\begin{abstract}
We discuss a novel pertubative QCD approach on the exclusive
non-leptonic two body B-meson decays.
We briefly review its ingredients and some important theoretical
issues on the factorization approaches. We show numerical results
which is compatible with recent experimantal data for the charmless
B-meson decays. 
\end{abstract}

\date{\today}

\maketitle

\section{Introduction}

The aim of the study on weak decay in B-meson is two folds:
(1) To determine precisely the elements of CKM matrix and 
to explore the origin of CP-violation in low energy scale,
(2) To understand strong interaction physics related to the confinements
of quarks and gluons within hadrons.

Both tasks complement each other. An understanding
of the connection between quarks and hadron properties is 
a necessary prerequeste for a precise determination of CKM matrix
elements and CP-violating phases, so called KM-phase\cite{KM}. 

The theoretical description of hadronic weak decays is difficult
since nonperturbative QCD interactions are involved. This makes  
a difficult to interpret correctly data from asymmetric B-factories
and to seek the origin of CP violation. In the case of B-meson decays
into two light mesons, we can explain roughly branching ratios by 
using the factorization approximation \cite{BSW:85,BSW:87}.
Since B-meson is quite heavy, when it decays into two light mesons,
the final-state mesons are moving so fast that it is difficult to
exchange gluons between final-state mesons. Therefore the
amplitude can be written in terms of the product of weak decay constant 
and transition form factors 
by the factorization (color-transparancy) argument. 
In this approach we can not calculate non-factorizable
contributions and annihilation contributions even though which is 
not dominant. Because of this weakness, 
violation of CP symmetry can not be predicted correctly.

Recently two different QCD
approaches beyond naive and general factorization assumption
\cite{BSW:85,BSW:87,Ali:98,Cheng:99} 
was proposed:
(1) QCD-factorization in heavy quark limit \cite{BBNS:99,BBNS:00}
 in which non-factorizable terms and $a_{i}$ are calculable in some cases. 
(2) A Novel PQCD approach \cite{KLS:01,KLS:02,KLS:03} including 
the resummation effects of the transverse momentum carried by partons
inside meson.
In this talk, I discuss some important theoretical issues in the PQCD
factorization and numerical results for charmless B-decays.
   
\section{Ingredients of PQCD}
\paragraph{Factorization in PQCD}
The idea of pertubative QCD is as follows:
When heavy B-meson decays into two light mesons, the hard process is
dominant. Since two light mesons fly so fast with large momentum,
it is reasonable assumptions that the final-state interaction is not
important for charmless B-decays and hard gluons are needed to boost
the resting spectator quark to get large momentum and finally 
to hadronize a fast moving final meson. 
So the dominant process is that one hard gluon
is exchanged between specator quark and other four quarks.

Let's start with the lowest-order diagram 
of $B \to K\pi$. The soft divergences in the $B \to \pi$ form factor
can be factorized into a light-cone B meson wave function,
and the collinear divergences can be factorized into 
a pion distribution amplitude.
The finite pieces of them is absorbed into the hard part.
Then in the natural way we can factorize amplitude into two pieces:
$G \equiv H(Q,\mu) \otimes \Phi(m, \mu)$ where H stands for hard part
which is calculable with a perturbative way, and $\Phi$ is wave functions
which belong to the non-perturbative physics.

PQCD adopt the three scale factorization theorem \cite{Li:01}
based on the perturbative QCD formalism by Brodsky and Lepage \cite{BL},
and Botts and Sterman \cite{BS}, with the inclusion of the transverse
momentum components which was carried by partons inside meson.

We have three different scales: electroweak scale: $M_W$,
hard interaction scale: $t \sim O( \sqrt( \bar{\Lambda}m_b))$, 
and the factorization scale: $1/b$ where
$b$ is the conjugate variable of parton transverse momenta.
The dynamics below $1/b$ is completely non-perturbative and 
can be parameterized into meson wave funtions which is universal and
process independent. In our analysis
we use the results of light-cone distribution amplitudes (LCDAs)
by Ball \cite{PB:01,PB:02} with light-cone sum rule.

The ampltitude in PQCD is expressed as 
\begin{eqnarray}
A \,\, \sim \,\, C(t) \,\,\times \,\, H(t) \,\,\times \,\, \Phi(x) \,\, 
\times\,\, 
\exp\left[ -s(P,b) - 2 \, \myint_{1/b}^{t} \,\,
{d\mu \over \mu} \,\, \gamma_q(\alpha_s(\mu))
\right]  
\end{eqnarray}
where $C(t)$ are Wilson coefficients, $\Phi(x)$ are meson LCDAs
and variable $t$ is the factorized scale in hard part.   

\paragraph{Sudakov Suppression Effects}
When we include $k_{\perp}$,
the double logarithms $\ln^2(Pb)$ are generated 
from the overlap of collinear and soft divergence in radiative corrections
to meson wave functions,
where P is the dominant light-cone component of a meson momentum. 
The resummation of these double logarithms leads to a Sudakov form factor
$exp[-s(P,b)]$ in Eq.(1), which suppresses the long distance contributions
in the large $b$ region, and vanishes as $b > 1/\Lambda_{QCD}$.

This suppression renders $k_{\perp}^2$ flowing into the hard amplitudes
of order
\begin{eqnarray}
k_{\perp}^2\sim O(\bar\Lambda M_B)\;.
\end{eqnarray}
The off-shellness of internal particles then remain of
$O(\bar\Lambda M_B)$ even in the end-point region, and the singularities
are removed. This mechanism is so-called Sudakov suppression.

Du {\it et al.} have studied the Sudakov effects in the evaluation
of nonfactorizable amplitudes \cite{Du}. If equating these amplitudes
with Sudakov suppression included to the parametrization in QCDF, it was
observed that the corresponding cutoffs are located in the reasonable
range proposed by Beneke {\it et al.} \cite{BBNS:00}.
Sachrajda {\it et al.} have expressed an opposite opinion on the effect
of Sudakov suppression in \cite{GS}. However, their conclusion was drawn
based on a very sharp $B$ meson wave function, which is not favored by
experimental data.

Here I would like to commnent on the negative opinions on the large
$k_{\perp}^2 \sim O(\bar{\Lambda}M_B)$.
It is easy to understand the increase of $k_{\perp}^2$ from $O(\bar\Lambda^2)$,
carried by the valence quarks which just come out of the initial meson
wave functions, to $O(\bar\Lambda M_B)$, carried by the quarks which are
involved in the hard weak decays. Consider the simple deeply inelastic
scattering of a hadron. The transverse momentum $k_{\perp}$ carried by a
parton, which just come out of the hadron distribution function, is
initially small. After infinite many gluon radiations, $k_{\perp}$ becomes of
$O(Q)$, when the parton is scattered by the highly virtual photon,
where $Q$ is the large momentum transfer from the photon. The evolution
of the hadron distribution function from the low scale to $Q$ is described
by the Dokshitzer-Gribov-Lipatov-Altarelli-Parisi (DGLAP) equation
\cite{GL,AP}. The mechanism of the DGLAP evolution in DIS is similar to that
of the Sudakov evolution in exclusive $B$ meson decays. The difference
is only that the former is the consequence of the single-logarithm
resummation, while the latter is the consequence of the double-logarithm
resummation.
\begin{figure}[t]
  \resizebox{20pc}{!}{\includegraphics[height=1.0\textheight]{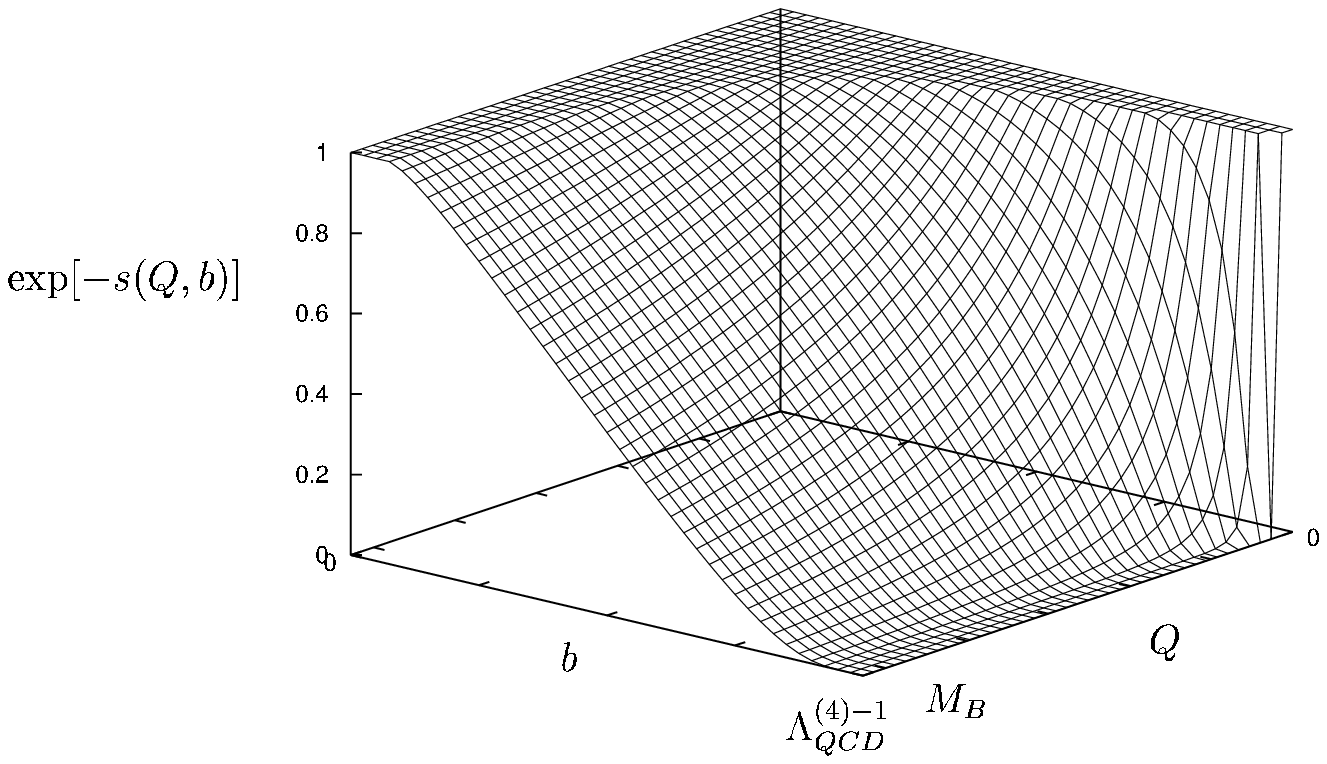}}
  \resizebox{15pc}{!}{\includegraphics[height=0.8\textheight]{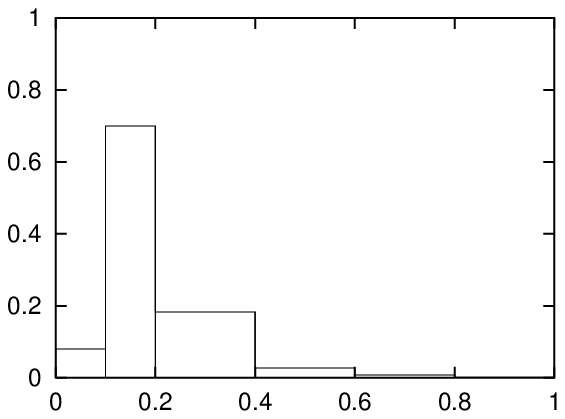}}
  \begin{picture}(0,0)(0,0)
   \put(-50,-5){${\alpha_s(t)}/{\pi}$}
   \put(-190,80){\rotatebox{90}{Fraction}}
  \end{picture}
  \caption{(a)Sudakov suppression factor (b)Fractional contribution to
  the $B \to \pi$ transition form factor $F^{B\pi}$ as a function of
$\alpha_s(t)/\pi$.}
\end{figure}

By including Sudakov effects, all contributions of the $B \to \pi$
form factor comes from the region with $\alpha_s/\pi < 0.3$ \cite{KLS:02}
as shown in Figure 1. It indicate that our PQCD results are well within
the perturbative region. 

\paragraph{Threshold Resummation}
The other double logarithm is $\alpha_s \ln^2(1/x)$ from the end point region
of the momentum fraction $x$ \cite{Li:02}. This double logarithm  is generated
by the corrections of the hard part in Figure 2.
This double logarithm can be factored out of the hard amplitude
systematically, and its resummation introduces a Sudakov factor 
$S_t(x)=1.78 [x(1-x)]^c$ with $c=0.3$ into PQCD factorization formula.
The Sudakov factor from threshold resummation 
is universal, independent of flavors of internal quarks, twists and topologies 
of hard amplitudes, and decay modes.
\begin{center}
\vspace{-20pt} \hfill \\
\begin{picture}
(70,0)(70,25)
\ArrowLine(80,10)(30,10)
\ArrowLine(130,10)(80,10)
\ArrowLine(30,-30)(80,-30)
\ArrowLine(80,-30)(130,-30)
\Gluon(50,10)(50,-30){3}{6}
\GBoxc(80,10)(7,7){0}
\GlueArc(80,10)(15,180,360){3}{6}
\end{picture}\hspace{15mm}
\begin{picture}
(70,0)(70,25)
\ArrowLine(80,10)(30,10)
\ArrowLine(130,10)(80,10)
\ArrowLine(30,-30)(80,-30)
\ArrowLine(80,-30)(130,-30)
\Gluon(110,10)(110,-30){3}{6}
\GBoxc(80,10)(7,7){0}
\GlueArc(80,10)(15,180,360){3}{6}
\end{picture} 
\end{center}

\vskip2.0cm
\begin{figure}[h]
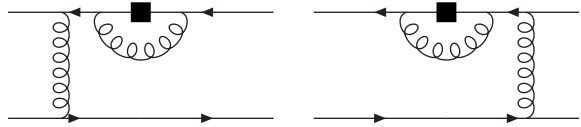

\caption{The diagrams generate double logarithm corrections 
for the threshold resummation.}
\end{figure}
\vskip1.0cm
Threshold resummation\cite{Li:02} and $k_{\perp}$ resummation 
\cite{CS,BS,StLi} 
arise from different
subprocesses in PQCD factorization and suppresses the
end-point contributions, making PQCD evaluation of exclusive $B$ meson
decays reliable. If excluding resummation effects, the PQCD predictions
for the $B\to K$ form factors are infrared divergent. If including only
$k_{\perp}$ resummation, the PQCD predictions are finite. However, the
two-parton twist-3 contributions are still huge, so that the $B\to K$
form factors have an unreasonably large value $F^{BK}\sim 0.57$ at maximal
recoil. The reason is that the double logarithms $\alpha_s\ln^2 x$ have
not been organized. If including both resummations, we obtain the
reasonable result $F^{BK}\sim 0.35$. These studies indicate the importance
of resummations in PQCD analyses of $B$ meson decays. In conclusion, if
the PQCD analysis of the heavy-to-light form factors is performed
self-consistently, there exist no end-point singularities, and both
twist-2 and twist-3 contributions are well-behaved.

\paragraph{Power Counting Rule in PQCD}
The power behaviors of various topologies of diagrams for two-body
nonleptonic $B$ meson decays with the Sudakov effects taken into account
has been discussed in details in \cite{CKL}. The relative importance is
summarized below:
\begin{eqnarray}
{\rm emission} : {\rm annihilation} : {\rm nonfactorizable} 
=1 : \frac{2m_0}{M_B} : \frac{\bar\Lambda}{M_B}\;,
\label{eq1}
\end{eqnarray}
with $m_0$ being the chiral symmetry breaking scale. The scale $m_0$
appears because the annihilation contributions are dominated by those
from the $(V-A)(V+A)$ penguin operators, which survive under helicity
suppression. In the heavy quark limit the annihilation and
nonfactorizable amplitudes are indeed power-suppressed compared to the
factorizable emission ones. Therefore, the PQCD formalism for two-body
charmless nonleptonic $B$ meson decays coincides with the factorization
approach as $M_B\to\infty$. However, for the physical value $M_B\sim 5$
GeV, the annihilation contributions are essential.
In Table 1 we can easily check the relative size of the different topology
in Eq.(\ref{eq1}) by the peguin contribution for W-emission
($f_{\pi}F^{P}$), annihilation($f_BF^{P}_a$) and
non-factorizable($M^P$) contributions.

Note that all the above topologies are of the same order in $\alpha_s$
in PQCD. The nonfactorizable amplitudes are down by a power of $1/m_b$,
because of the cancellation between a pair of nonfactorizable diagrams,
though each of them is of the same power as the factorizable one. I
emphasize that it is more appropriate to include the nonfactorizable
contributions in a complete formalism. The
factorizable internal-$W$ emisson contributions are strongly suppressed
by the vanishing Wilson coefficient $a_2$ in the $B\to J/\psi K^{(*)}$
decays \cite{YL}, so that nonfactorizable contributions become dominant.
In the $B\to D\pi$ decays, there is no soft cancellation between a pair
of nonfactorizable diagrams, and nonfactorizable contributions are
significant \cite{YL}.

In QCDF the factorizable and nonfactorizable amplitudes are of the same
power in $1/m_b$, but the latter is of next-to-leading order in
$\alpha_s$ compared to the former. Hence, QCDF approaches FA in the
heavy quark limit in the sense of $\alpha_s\to 0$. Briefly speaking,
QCDF and PQCD have different counting rules both in $\alpha_s$ and in
$1/m_b$. The former approaches FA logarithmically
($\alpha_s\propto 1/\ln m_b \to 0$), while the latter does linearly
($1/m_b\to 0$).
\begin{table}[t]
\begin{tabular}{|c|cc|c|} \hline 
Amplitudes & Left-handed gluon exchange & 
Right-handed gluon exchange & Total \\
\hline 
$Re(f_{\pi} F^T)$ & \hspace{0.5cm}$7.07 \cdot 10^{-2}$ & 
\hspace{0.5cm}$3.16 \cdot 10^{-2}$
 & $1.02 \cdot 10^{-1}$   \\
$Im(f_{\pi} F^T)$ & $-$  & $-$ &  $-$ \\ 
\hline
$Re(f_{\pi} F^P)$ &  \hspace{0.5cm}-$5.52 \cdot 10^{-3}$ & 
\hspace{0.5cm}-$2.44 \cdot 10^{-3}$ 
& -$7.96 \cdot 10^{-3}$ \\
$Im(f_{\pi} F^P)$ & $-$ & $-$ & $-$ \\
\hline
$Re(f_{B} F_a^P)$ & \hspace{0.5cm}$4.13 \cdot 10^{-4}$ & 
\hspace{0.5cm}-$6.51 \cdot 10^{-4}$
& -$2.38 \cdot 10^{-4}$  \\
$Im(f_{B} F_a^P)$ & \hspace{0.5cm}$2.73 \cdot 10^{-3}$ & 
\hspace{0.5cm}$1.68 \cdot 10^{-3}$ 
& $4.41 \cdot 10^{-3}$ \\
\hline 
$Re(M^T)$ & \hspace{0.5cm}$7.06 \cdot 10^{-3}$ & 
\hspace{0.5cm}-$7.17 \cdot 10^{-3}$ 
& -$1.11 \cdot 10^{-4}$ \\
$Im(M^T)$ & \hspace{0.5cm}-$1.10 \cdot 10^{-2}$ & 
\hspace{0.5cm}$1.35 \cdot 10^{-2}$
& $2.59 \cdot 10^{-3}$  \\
\hline
$Re(M^P)$ & \hspace{0.5cm}-$3.05 \cdot 10^{-4}$ & 
\hspace{0.5cm}$3.07 \cdot 10^{-4}$
& $2.17 \cdot 10^{-6}$  \\
$Im(M^P)$ & \hspace{0.5cm}$4.50 \cdot 10^{-4}$ & 
\hspace{0.5cm}-$5.29 \cdot 10^{-4}$
& -$7.92 \cdot 10^{-5}$ \\
\hline
$Re(M_a^P)$ & \hspace{0.5cm}$2.03 \cdot 10^{-5}$ & 
\hspace{0.5cm}-$1.37 \cdot 10^{-4}$
& -$1.16 \cdot 10^{-4}$ \\
$Im(M_a^P)$ & \hspace{0.5cm}-$1.45 \cdot 10^{-5}$ & 
\hspace{0.5cm}-$1.27 \cdot 10^{-4}$ 
& -$1.42 \cdot 10^{-4}$ \\ \hline
\end{tabular}
\label{TABLE11.3}
\caption{Amplitudes for the $B_d^{0} \to K^{+} \pi^{-}$ decay 
where $F$ ($M$) denotes factorizable (nonfactorizable) 
contributions, $P$ ($T$) denotes the penguin (tree) contributions,
and $a$ denotes the annihilation contributions. Here we adopted 
$\phi_3=80^0$ and $R_b=\sqrt(\rho^2 +\eta^2)=0.38$.  }
\end{table} 
\section{Important Theoretical Issues}
\paragraph{End Point Singularity and Form Factors}
If calculating the $B\to\pi$ form factor $F^{B\pi}$ at large recoil using
the Brodsky-Lepage formalism \cite{BL,BSH}, a difficulty immediately
occurs. The lowest-order diagram for the hard amplitude is proportional to 
$1/(x_1 x_3^2)$, $x_1$ being the momentum fraction associated with the
spectator quark on the $B$ meson side. If the pion distribution amplitude
vanishes like $x_3$ as $x_3\to 0$ (in the leading-twist, {\it i.e.},
twist-2 case), $F^{B\pi}$ is logarithmically divergent. If the pion
distribution amplitude is a constant as $x_3\to 0$ (in the
next-to-leading-twist, {\it i.e.}, twist-3 case), $F^{B\pi}$ even becomes
linearly divergent. These end-point singularities have also appeared in
the evaluation of the nonfactorizable and annihilation amplitudes in QCDF
mentioned above.

When we include small parton transverse momenta $k_{\perp}$, we have
\begin{equation}
{1 \over x_1\,\, x_3^2 M_B^4} \hspace{10mm} \rightarrow
\hspace{10mm} {1 \over (x_3\, M_B^2 + k_{3\perp}^2) \,\,
[x_1x_3\, M_B^2 + (k_{1\perp} - k_{3\perp})^2]}
\label{eq:4} 
\end{equation}
and the end-point singularity is smeared out.

In PQCD, we can calculate analytically space-like form factors for $B \to P,V$
transition and
also time-like form factors for the annihilation process \cite{CKL,Kurimoto}.

\paragraph{Strong phases}
While stong phases in FA and QCDF 
come from the Bander-Silverman-Soni (BSS) mechanism\cite{BSS}
and from the final state interaction (FSI), the dominant strong phase in PQCD
come from the factorized annihilation diagram\cite{KLS:01,KLS:02,KLS:03}.
It has been argued that the two sources of strong phases in the FA
and QCDF approaches are in fact strongly suppressed by the charm mass
threshold and by the end-point behavior of meson wave functions.

\begin{table}[t]
\begin{tabular}{|c|ccc|c|c|} \hline 
Decay Channel & CLEO & BELLE & BABAR & World Av. & PQCD  \\
\hline  
$\pi^{+}\pi^{-}$ & $4.3^{+1.6}_{-1.4}\pm 0.5$ &
 $5.6^{+2.3}_{-2.0}\pm 0.4$ &
 $4.1\pm 1.0 \pm 0.7$ &  
$4.4 \pm 0.9$ &
$7.0^{+2.0}_{-1.5}$  \\
$\pi^{+}\pi^{0}$ & $5.6^{+2.6}_{-2.3}\pm1.7$ & 
 $7.8^{+3.8+0.8}_{-3.2-1.2}$ &
 $5.1^{+2.0}_{-1.8} \pm 0.8$ & $5.6\pm1.5$ & 
$3.7^{+1.3}_{-1.1}$    \\ 
$\pi^{0}\pi^{0}$ & $<5.7$ & 
 $-$ &  $-$ & $-$ &
  $0.3 \pm 0.1$    \\ 
\hline
$K^{0}\pi^{\pm}$ & 
 $18.2^{+4.6}_{-4.0}\pm 1.6$ &
 $13.7^{+5.7 +1.9}_{-4.8 -1.8}$ &  
 $18.2^{+3.3}_{-3.0}\pm 2.0$ &
 $17.3 \pm 2.7$ &  
$16.4^{+3.3}_{-2.7}$  \\
$K^{\pm}\pi^{\mp}$ & 
 $17.2^{+2.5}_{-2.4}\pm 1.2$ &
 $19.3^{+3.4+1.5}_{-3.2-0.6}$ &  
 $16.7\pm 1.6 \pm 1.3$ &
 $17.3 \pm 1.5$ &  
 $15.5^{+3.1}_{-2.5}$    \\ 
$K^{\pm}\pi^{0}$ &
 $11.6^{+3.0+1.4}_{-2.7-1.3}$ &
 $16.3^{+3.5+1.6}_{-3.3-1.8}$ &  
 $10.8^{+2.1}_{-1.9} \pm 1.0$ &
 $12.1 \pm 1.7$ &  
  $9.1^{+1.9}_{-1.5}$    \\
$K^{0}\pi^{0}$ &
 $14.6^{+5.9+2.4}_{-5.1-3.3}$ &
 $16.0^{+7.2+2.5}_{-5.9-2.7}$ &  
 $8.2^{+3.1}_{-2.7} \pm 1.2$ &
 $10.4 \pm 2.7$ & 
 $8.6 \pm 0.3$    \\ 
\hline 
\end{tabular}
\label{TABLE11.2}
\caption{Branching ratios of $B \to \pi \pi $ and $K \pi $ decays 
with $\phi_3=80^0$, $R_b=0.38$. Here we adopted
$m_0^{\pi}=1.3$ GeV and $m_0^{K}=1.7$ GeV. $R_b=\sqrt(\rho^2+\eta^2)$ and
unit is $10^{-6}$. (Aug/2001 data)}
\end{table} 
\paragraph{Dynamical Penguin Enhancement vs Chiral Enhancement}
As explained before, the hard scale is about 1.5 GeV.
Since the RG evolution of the Wilson coefficients $C_{4,6}(t)$ increase
drastically as $t < M_B/2$, while that of $C_{1,2}(t)$ remain almost
constant, we can get a large enhancement effects from both wilson
coefficents and matrix elements in PQCD. 
 
In general the amplitude can be expressed as
\begin{equation}
Amp \sim [a_{1,2} \,\, \pm \,\, a_4 \,\,
\pm \,\, m_0^{P,V}(\mu) a_6] \,\, \cdot \,\, <K\pi|O|B>
\label{eq:2}
\end{equation}
with the chiral factors $m_0^P(\mu)=m_P^2/[m_1(\mu)+m_2(\mu)]$ for
pseudoscalr meson 
and $m_0^{V}= m_V$ for vector meson.
To accommodate the $B\to K\pi$ data in the factorization and
QCD-factorization approaches, one relies on the chiral enhancement by
increasing the mass $m_0$ to as large values about 3 GeV at $\mu=m_b$ scale.
So two methods accomodate large branching ratios of $B \to K\pi$ and
it is difficult for us to distinguish two different methods in $B \to
PP$ decays. However we can do it in $B \to PV$ because there is no
chiral factor in LCDAs of the vector meson. 

We can test whether dynamical enhancement or chiral enhancement is responsible
for the large $B \to K\pi$ branching ratios by measuring the $B \to \phi K$ modes.
In these modes penguin contributions dominate, such that their branching ratios are
insensitive to the variation of the unitarity angle $\phi_3$.
According to recent
works by Cheng {\it at al.} \cite{CK}, 
the branching ratio of $B \to \phi K$ is $(2-7)
\times 10^{-6}$ including $30\%$ annihilation contributions in
QCD-factorization approach. 
However PQCD predicts $10 \times 10^{-6}$ \cite{CKL,Mishima}.   

\paragraph{Fat Imaginary Penguin in Annihilation}
There is a falklore that annihilation contribution is negligible
compared to W-emission one. In this reason annihilation contribution
was not included in the general factorization approach and the first
paper on QCD-factorization by Beneke et al. \cite{BBNS:99}.
In fact there is a suppression effect for the operators with structure
$(V-A)(V-A)$ because of a mechanism similar to the helicity
suppression for $\pi \to \mu \nu_{\mu}$. However annihilation from 
the operators $O_{5,6,7,8}$ with the structure $(S-P)(S+P)$ via Fiertz
transformation survive under the helicity suppression and can get
large imaginary value. The real part of factorized annihilation contribution
becomes small because there is a cancellation between left-handed
gluon exchanged one and right-handed gluon exchanged one as shown in
Table 1. This mostly pure imaginary value of annihilation is a main
source of large CP asymmetry in $B \to \pi^{+}\pi^{-}$ and $K^{+}\pi^{-}$.
In Table 5 we summarize the CP asymmetry in 
$B \to K(\pi)\pi$ decays.

\begin{table}[t]
\begin{tabular}{|c|ccc|c|} \hline
Decay Channel & CLEO & BELLE & BABAR & PQCD   \\
\hline  
$\phi K^{\pm}$ & 
 $5.5^{+2.1}_{-1.8}\pm 0.6$ &
 $11.2^{+2.2}_{-2.0} \pm 0.14$ &  
 $7.7^{+1.6}_{-1.4}\pm 0.8$ & 
 $10.2^{+3.9}_{-2.1}$  \\
$\phi K^{0}$ & 
 $ < 12.3 $ &
 $8.9^{+3.4}_{-2.7}\pm 1.0$ &  
 $8.1^{+3.1}_{-2.5}\pm 0.8 $ &
 $9.6^{+3.7}_{-2.0}$    \\ 
\hline
$K^{*0} \pi^{\pm}$ & 
 $7.6^{+3.5}_{-3.0} \pm 1.6$ &
 $19.4^{+4.2}_{-3.9} \pm 2.1^{+3.5}_{-6.8}$ &  
 $15.5 \pm 3.4 \pm 1.5$ & 
 $12.2^{+2.4}_{-2.0} $  \\
$K^{*\pm}\pi^{\mp}$ & 
 $22^{+8+4}_{-6-5} $ &
 $-$ &  
 $-$ &
 $9.6^{+2.0}_{-1.6}$    \\ 
\hline 
\end{tabular}
\label{TABLE11.7}
\caption{Branching ratios of $B \to \phi K^{(*)} $ decays 
with $\phi_3=80^0$, $R_b=0.38$. Here we adopted $m_0^{\pi}=1.3$ GeV
and $m_0^{K}=1.7$ GeV. $R_b=\sqrt(\rho^2 + \eta^2)$ and
unit is $10^{-6}$. (Aug/2001 data)}
\end{table} 
\section{Numerical Results}
\paragraph{ Branching ratios and Ratios of CP-averaged rates}
The PQCD approach allows us to calculate the amplitudes for charmless B-meson decays
in terms of ligh-cone distribution amplitudes upto twist-3. We focus on decays
whose branching ratios have already been measured. 
We take allowed ranges of shape parameter for the B-meson wave funtion as 
$\omega_B = 0.36-0.44$ which accomodate to reasonable form factors, 
$F^{B\pi}(0)=0.27-0.33$ and $F^{BK}(0)=0.31-0.40$. We use values of chiral factor
with $m_0^{\pi}=1.3 GeV$ and $m_0^{K}=1.7 GeV$.
Finally we obtain branching ratios for $B\to K(\pi)\pi$ \cite{CKL,LUY}, 
$K\phi$ \cite{CKL,Mishima} and $K^{*}\pi$,
which is well agreed with present experimental data (see Table 2 and 3).

In order to reduce theoretical uncertainties from decay constant of B-meson
and from light-cone distribution amplitudes, we consider rates of CP-averaged
branching ratios, which is presented in Table 4.
While the first ratio is hard to be explained by QCD factorization approach
with $\phi_3 < 90^o$, our prediction can be reached to 0.30. 

\paragraph{CP Asymmetry of $B \to \pi\pi, K\pi$}
Because we have a large imaginary contribution from factorized 
annihilation diagrams in PQCD approach,
we predict large CP asymmetry ($\sim 20 \%$) in $B^0 \to \pi^{+}\pi^{-}$ decays
and about $-15 \%$ CP violation effects in  $B^0 \to K^{+}\pi^{-}$.
The detail prediction is given in Table 5.
The precise measurement of direct CP asymmetry (both magnitude and sign) 
is a crucial way to test factorization models 
which have different sources of strong phases.
Our predictions for CP-asymmetry on $B\to K(\pi)\pi$ have a totally opposite
sign to those of QCD factorization.

\begin{table}[t]
\begin{tabular}{|c||c||c|c|} \hline 
~~~~~~Quatity~~~~~~ & ~~~~~~CLEO~~~~~~ & ~~~~~~PQCD~~~~~~  
& ~~~~~~BBNS~~~~~~  \\ \hline 
  &   &   &   \\
${Br(\pi^{+} \pi^{-}) \over Br(\pi^{\pm} K^{\mp}) }$ & $0.25 \pm 0.10$ & 
 $0.30-0.69$ & $0.5-1.9$  \\
  &   &   &   \\
  &   &   &   \\
${Br(\pi^{\pm} K^{\mp}) \over 2 Br(\pi^{0} K^{0}) }$ & $0.59 \pm 0.27$ & 
 $0.78-1.05$ & $0.9-1.4$  \\
  &   &   &   \\
  &   &   &   \\
${2 \,\, Br(\pi^{0} K^{\pm}) \over Br(\pi^{\pm} K^{0}) }$ & $1.27 \pm 0.47$ & 
 $0.77-1.60$ & $0.9-1.3$  \\
  &   &   &   \\
  &   &   &   \\
${\tau(B^{+}) \over \tau(B^0)} \,
{Br(\pi^{\mp} K^{\pm}) \over Br(\pi^{\pm} K^{0}) }$ & $1.00 \pm 0.30$ & 
 $0.70-1.45$ & $0.6-1.0$  \\
  &   &   &   \\
\hline
\end{tabular}
\label{TABLE11.5}
\caption{Ratios of CP-averaged rates in $B \to K \pi, \pi\pi $ decays 
with $\phi_3=80^0$, $R_b=0.38$. Here we adopted
$m_0^{\pi}=1.3$ GeV, $m_0^{K}=1.7$ GeV and $R_b=\sqrt(\rho^2 + \eta^2)$.
(Aug/2001 data) }
\end{table} 

\begin{table}[htb]
\begin{tabular}{|c|c|ccc|} \hline 
~~~~~~$A_{CP}(\%)$~~~~~~ & ~~~~~~Experiment~~~~~~ & ~~~& Theory &~~~~~~ \\ 
  &   
& ~~~~PQCD~~~~ & ~~~~BBNS\cite{BBNS:00}~~~~ & ~~~~CFMPS\cite{charmpenguin}~~ ($|A_{CP}|$) \\
\hline 
  &   &   &   & \\
$\pi^{+} K^{-}$ & $-4.8 \pm 6.8$ & $-12.9 \sim -21.9$ & $5\pm9$ & $17\pm6$  \\
  & (BaBar) \, $-7\pm8\pm2$   &  &  & \\
  &   &  &  &  \\
$\pi^{0}K^{-}$ & $-9.6 \pm 11.9$ & 
 $-10.0 \sim -17.3$ & $7\pm9$ & $18\pm6$  \\
  &   &   &  &  \\
  &   &   &  & \\
$\pi^{-}\bar{K}^{0}$ & $-4.7 \pm 13.9$ & 
 $-0.6 \sim -1.5$ & $1\pm1$ & $3\pm3$  \\
  &   &   &  &  \\
  &   &   &  &  \\
$\pi^{+}\pi^{-}$ & $-25 \pm 48$ & 
 $15.0 \sim 30.0$ & $-6\pm12$ & $58\pm29$ \\
  &   &   &   &  \\
\hline
\end{tabular}
\label{TABLE11.6}
\caption{CP-asymmetry in $B \to K \pi, \pi\pi $ decays 
with $\phi_3=40^0 \sim 90^0$, $R_b=0.38$. Here we adopted
$m_0^{\pi}=1.3$ GeV, $m_0^{K}=1.7$ GeV and $R_b=\sqrt(\rho^2 + \eta^2)$.
(Aug/2001 data)}
\end{table} 

\paragraph{Determination $\phi_3$ in $B \to \pi\pi, K\pi$}
Some years ago, many authors\cite{gamma:01, gamma:02, gamma:03} have derived
a bound on $\phi_3$ from rates of branching ratios on $B \to K(\pi) \pi$.
The ratio  $R_{K} $ is given by
\begin{equation}
R_{K} \equiv {Br(B^0\to K^{\pm}\pi^{\mp}) \over Br(B^{\pm}\to K^{0}\pi^{\pm})}
= 1.0 + {2 \,\, \lambda^2 \,\, R_b \over a_K} \,\, cos\phi_3
\end{equation}
where $\lambda=0.22$, 
$R_b=0.41\pm 0.07$ and $a_K = (a_4 + 2\, a_6 \, r_K)/a_1$.
As shown in Figure 3, arbitrary $\phi_3$ is allowed due to 
large experimantal uncertainties in present data ($0.95\pm0.30$).
We expect more precise measurement of $R_K$ to determine $\phi_3$ 
in future within 2-3 years. 
\begin{figure}[t]
\caption{Dependence of the ratio $R_K$ on $\phi_3$.
The dashed (dotted) lines correspond to the bounds (central value)
of the data.}
{\rotatebox{-90}{\includegraphics[height=.5\textheight]{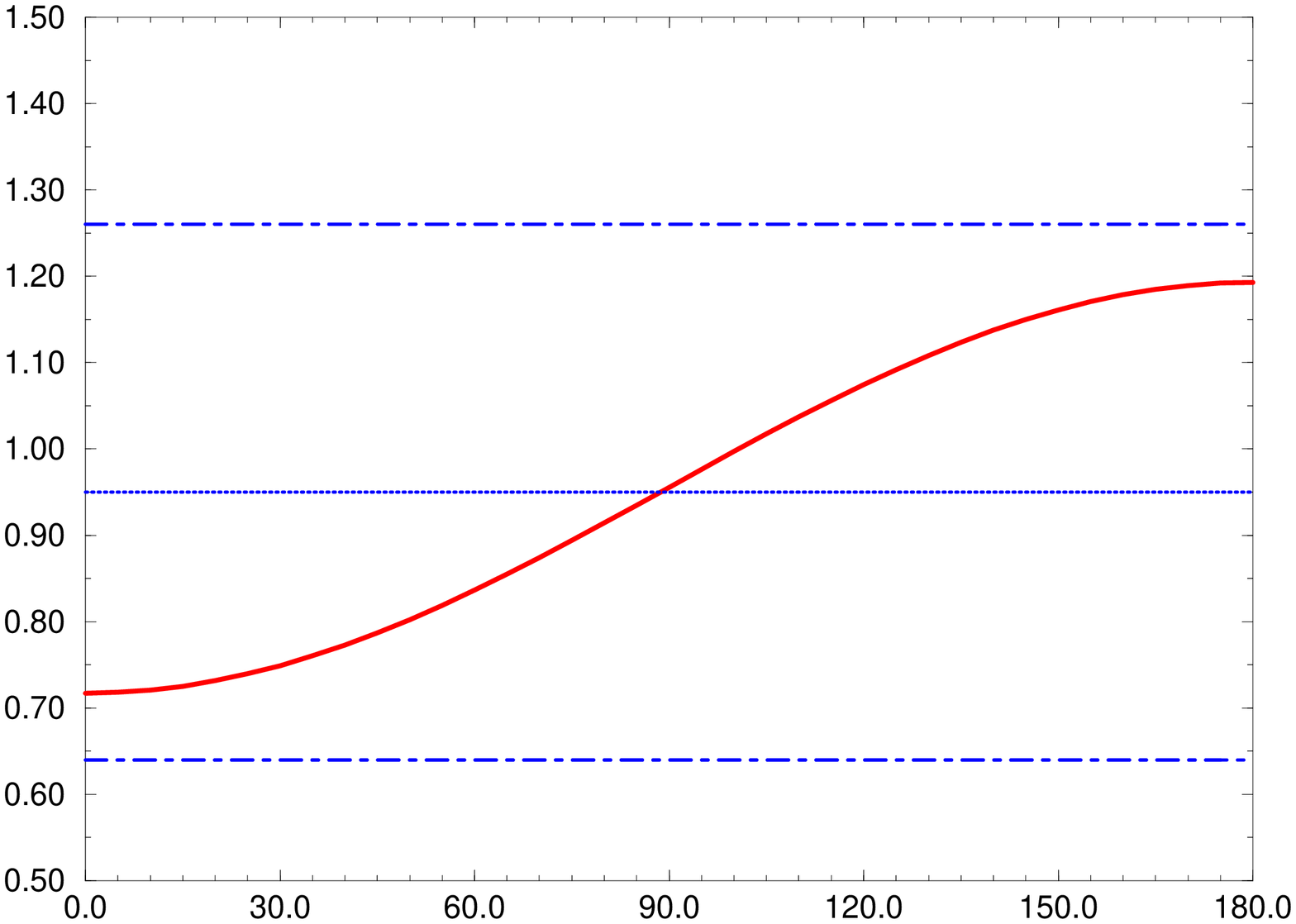}}}
  \begin{picture}(0,0)(0,0)
   \put(-180,-225){${\phi_3(degree)}$}
   \put(-300,-130){\rotatebox{90}{$R_K$}}
  \end{picture}
\end{figure}

\section{Summary and Outlook}
In this paper I have discussed ingredients of PQCD approach and some important
theoretical issues with numerical results by comparing exparimental data.
A new PQCD factorization approach provides a useful theoretical framework
for a systematic analysis on non-leptonic two-body B-meson decays.
This approach explain sucessfully present experimental data upto now
and will be tested more thoroughly with more precise data in near future. 

\begin{theacknowledgments}
We wish to acknowlege joyful discussions with S.J. Brodsky 
and other members of PQCD working group.
This work was supported in part by Grant-in Aid of Special
Project Research (Physics of CP Violation) and
by Grant-in Aid for Scientific Exchange from the Ministry of Education, 
Science and Culture of Japan.
Y.Y.K. thanks H.Y. Cheng and M. Kobayashi for their hospitality. 
and encouragement.
\end{theacknowledgments}



\begin{thebibliography}{99}

\bibitem{KM}
Kobayashi, M. and Maskawa,T., \emph{Prog. Theor. Phys.} {\bf 49}, 652 (1973).

\bibitem{BSW:85}
Bauer, M., Stech, B. and Wirbel, M., \emph{Z. Phys. {\bf C}} \textbf{29}, 637
  (1985).

\bibitem{BSW:87}
Bauer, M., Stech, B. and Wirbel, M., \emph{Z. Phys. {\bf C}} \textbf{34}, 103
  (1987). 

\bibitem{Ali:98}
Ali, A., Kramer, G. and Lu, C.-D., \emph{Phys. Rev. {\bf D}} \textbf{58},
  094009 (1998).

\bibitem{Cheng:99}
Chen, Y.-H., Cheng, H.-Y., Tseng, B. and Yang, K.-C., \emph{Phys. Rev. {\bf
  D}} \textbf{60}, 094014 (1999).

\bibitem{BBNS:99}
Beneke, M., Buchalla, G., Neubert, M. and Sachrajda, C.~T., \emph{Phys. Rev.
  Lett.} \textbf{83}, 1914 (1999).

\bibitem{BBNS:00}
Beneke, M., Buchalla, G., Neubert, M. and Sachrajda, C.~T., \emph{Nucl. Phys.
  {\bf B}} \textbf{591}, 313 (2000).

\bibitem{KLS:01}
Keum, Y.-Y., Li, H.-N. and Sanda, A.~I., \emph{Phys. Lett. {\bf B}}
  \textbf{504}, 6 (2001).

\bibitem{KLS:02}
Keum, Y.-Y., Li, H.-N. and Sanda, A.~I., \emph{Phys. Rev. {\bf D}}
  \textbf{63}, 074006 (2001).

\bibitem{KLS:03}
Keum, Y.-Y. and Li, H.-N., \emph{Phys. Rev. {\bf D}} \textbf{63}, 054008
  (2001).

\bibitem{Li:01}
Chang, C.~H. and Li, H.-N., \emph{Phys. Rev. {\bf D}} \textbf{55}, 5577
  (1997).

\bibitem{BL}
Lepage, G.~P. and Brodsky, S., 
\emph{Phys. Rev. {\bf D}} \textbf{22}, 2157  (1980).

\bibitem{BS}
Botts, J. and Sterman, G., \emph{Nucl. Phys. {\bf B}} \textbf{225}, 62
  (1989).

\bibitem{PB:01}
Ball, P., \emph{JHEP} \textbf{9809}, 005 (1998).

\bibitem{PB:02}
Ball, P., \emph{JHEP} \textbf{9901}, 010 (1999).

\bibitem{Du}
Du, D.-S., Huang, C.-S., Wei, Z.-T. and Yang, M.-Z., 
\emph{Phys. Lett. {\bf B}} \textbf{520}, 50 (2001).

\bibitem{GS}
Descotes-Genon, S. and Sachrajda, C.~T., \emph{hep-ph/0109260}.

\bibitem{GL}
Gribov, V.~N. and Lipatov, L.~N., 
\emph{Sov. J. Nucl. Phys.} \textbf{15}, 428 (1972).

\bibitem{AP}
Altarelli, G. and Parisi, G., \emph{Nucl. Phys. {\bf B}} \textbf{126}, 298
  (1977).

\bibitem{Li:02}
Li, H.-N., \emph{hep-ph/0102013}.

\bibitem{CS}
Collins, J.~C. and Soper, D.~E., \emph{Nucl. Phys. {\bf B}} \textbf{193}, 381
  (1981).

\bibitem{StLi}
Li, H.-N. and Sterman, G., \emph{Nucl. Phys. {\bf B}} \textbf{381}, 129
  (1992).

\bibitem{CKL}
Chen, C.-H., Keum, Y.-Y. and Li, H.-N., \emph{Phys. Rev. {\bf D}}
  \textbf{64}, 112002 (2001).

\bibitem{YL}
Yeh, T.-W. and Li, H.-N., \emph{Phys. Rev. {\bf D}} \textbf{56}, 1615 (1997).

\bibitem{BSH}
Szczepaniak, A., Henley, E.~M. and Brodsky, S., \emph{Phys. Lett. {\bf B}}
  \textbf{243}, 287 (1990).

\bibitem{Kurimoto}
Kurimoto, T., Li, H.-N. and Sanda, A.~I., \emph{Phys. Rev. {\bf D}}
  \textbf{65}, 014007 (2002).

\bibitem{BSS}
Bander, M., Silverman, D. and Soni, A., \emph{Phys. Rev. Lett.} \textbf{43},
  242 (1979).

\bibitem{CK}
Cheng, H.-Y. and Yang, K.-C., \emph{Phys. Rev. {\bf D}} \textbf{64}, 074004
  (2001).

\bibitem{Mishima}
Mishima, S., \emph{Phys. Lett. {\bf B}} \textbf{521}, 252 (2001).

\bibitem{LUY}
Lu, C.~D., Ukai, K. and Yang, M.~Z., \emph{Phys. Rev. {\bf D}} \textbf{63},
  074009 (2001).

\bibitem{charmpenguin} Ciuchini,~M., Franco,~E., Martinelli,~G., Pierini,~M.,
and Silverstrini,L., \emph{Phys. Lett. {\bf B}} \textbf{515}, 33 (2001).

\bibitem{gamma:01}
Mannel, T. and Fleischer, R., \emph{Phys. Rev. {\bf D}} \textbf{57}, 2752
  (1998).

\bibitem{gamma:02}
Buras, A.~J. and Fleischer, R., \emph{Euro. Phys. J. {\bf C}} \textbf{16}, 97
  (2000).

\bibitem{gamma:03}
Neubert, M. and Rosner, J.~L., \emph{Phys. Lett. {\bf B}} \textbf{441}, 403
  (1998).

\end{thebibliography}
\end{document}